\documentclass[global]{svjour}
\usepackage{amsfonts}
\usepackage{amsmath}

\input{tcilatex}

\begin{document}

\title{Statistical Mechanical Theory of a Closed Oscillating Universe}
\author{A. \ P\'{e}rez-Madrid\inst{1} \and I. Santamar\'{\i}a-Holek\inst{2}}
\institute{Dpt. de F\'{\i}sica Fonamental, Facultat de F\'{\i}sica, Universitat de
Barcelona. Av. Diagonal 647, 08028 Barcelona, Spain. 
\email{agustiperezmadrid@ub.edu}%
\and Facultad de Ciencias, Universidad Nacional Aut\'{o}noma de M\'{e}xico.
Circuito exterior de Ciudad Universitaria. 04510, D.F. M\'{e}xico.}
\maketitle

\begin{abstract}
Based on Newton's laws reformulated in the Hamiltonian dynamics combined
with statistical mechanics, we formulate a statistical mechanical theory
supporting the hypothesis of a closed oscillating universe. We find that the
behaviour of the universe as a whole can be represented by a free entropic
oscillator whose lifespan is nonhomogeneous, thus implying that\textbf{\ }%
time is shorter or longer according to the state of the universe itself
given through its entropy. We conclude that time reduces to the entropy
production of the universe and that a nonzero entropy production means that
local fluctuations could exist giving rise to the appearance of masses and
to the curvature of the space.
\end{abstract}

\keywords{nonequilibrium statistical mechanics-irreversibility- cosmological
theories}

\section{Introduction}

Cosmology, as an important discipline of scientific knowledge, tries to
address fundamental questions beyond the realm of sciences and influencing
our philosophical and even religious comprehension of the world. Perhaps the
major question posed by cosmology is the origin of our universe. Nowadays,
the most important theory providing an answer to this question is the
Big-Bang theory \cite{hawking}. Aside from others details, we are mainly
interested here in one of the implications of the Big-Bang theory, i. e.,
\textquotedblright that the universe will one day end\textquotedblright , 
\cite{hawking}. According to a current speculation, in a closed universe,
after the expansion eventually stops, a contraction will follow leading to
an implosion into a singularity a process known as the Big Crunch. This
oscillating behavior of the universe has important implications over the
concept of time, first because it is not clear that the initial and final
states of the universe can be correctly conceptualized through the notion of
time, understood as a fundamental quantity, and secondly because this
behavior suggests time is created and destroyed at the beginning and the end
of the universe.

In contrast with these implications, we know that in all fundamental
physical theories time plays a central role. Mechanics, quantum mechanics
and relativistic theories require time as a quantity through which the
evolution of the system is conceptualized. However, thermodynamics does not
require the concept of time in order to establish the \textquotedblright
direction\textquotedblright\ in which spontaneous natural processes occur,
that is, towards the state of thermodynamic equilibrium. These
considerations lead to the following question: Is the time a fundamental
quantity to describe the evolution of a system during a spontaneous natural
process?

Beside the Big Bang theory other alternatives with respect to the origin and
nature of the universe \ have been posed: is it open or closed...? it is
cyclic?, or is there an arrow of time related with entropy and
irreversibility?. Therefore these questions are still a recurrent topic of
debate \cite{matias}-\cite{steinhardt}.

Here, we present a cosmological statistical mechanics theory of a closed
universe \textbf{\ }not in equilibrium,\textbf{\ }oscillating between two
thermodynamic equilibrium states which runs parallel to\textbf{\ }other
cosmological models proposing an oscillating or cyclic universe \cite%
{lessner}, \cite{ruth}, \cite{steinhardt}. Our statistical mechanical theory
might complement these models.\textbf{\ } We assume that physical
interactions give rise to time and space, that is, we consider that if
physical interactions did not exist, time and space would not exist either.
Our theory provides us with an expression of the\textbf{\ }entropy
production of the universe based on first principles from wich one infers
that time is inhomogeneous and reduces to the production of entropy.
Moreover, our theory implies that there is no global time arrow although it
explains the existence of irreversible local phenomena \cite{agusti} \textbf{%
\ }as the corresponding ones in a nonequilibrium fluid. A more general
statistical theory can be elaborated by incorporating quantum and
relativistic effects through a description in terms of the set of reduced
density operators instead of the distribution vector in addition to the
generalized von Neumann equation and the generalized von Neumann entropy.
Density operators will account for quantum and also relativistic effects
since the metrics of the space-time will also be contained in this
description\cite{agusti2}. However, since the stochastic dynamic is
determined by the eigenvalues of the Hamiltonian, this more mathematically
elaborate theory will essentially lead to the same conclusions.

To mathematically formalize the previous statements, we will consider an
isolated system composed by N particles or bodies (the closed universe)
whose large-scale structure is governed by long-range interactions falling
into the Hamiltonian of the universe itself.

In classical mechanics the state of an N-body system at any time is given by
a set of 3N generalized coordinates $q_{1},......,q_{3N}$ and 3N conjugated
generalized momenta $p_{1},......,p_{3N}$. The value of these 6N variables
defines a point in the 6N-dimensional phase space $\mathbf{\Gamma }$
corresponding to the system. This representative point of the system moves
in the phase space along a trajectory determined by Hamilton's equations%
\begin{equation}
\overset{\cdot }{q}_{l}=\frac{\partial H}{\partial p_{l}}\text{ \ , }\overset%
{\cdot }{p}_{l}=-\frac{\partial H}{\partial q_{l}}\text{ \ ,}
\label{equ-mot}
\end{equation}%
where $H$ is the Hamiltonian of the universe. Therefore, Hamilton's
equations describe a flow in $\mathbf{\Gamma }$ space whose density flow $F(%
\mathbf{\Gamma },t)$ varies according to the Liouville equation%
\begin{equation}
i\frac{\partial }{\partial t}F\mathbf{=}LF  \label{liouville}
\end{equation}%
with%
\begin{equation}
L=i\left\{ H,..\right\} =i\sum_{l}\left( \frac{\partial H}{\partial q_{l}}%
\frac{\partial }{\partial p_{l}}\text{ }-\frac{\partial H}{\partial p_{l}}%
\text{\ }\frac{\partial }{\partial q_{l}}\right)  \label{liou-op}
\end{equation}%
being the Liouville operator. The description in terms of the Liouville
equation is completely equivalent to that in terms of Hamilton's equations 
\cite{balescu}. If the system occupies a volume $V$ in the configurational
space the density will vanish out of this volume, a fact which can be used
to show that the Liouville operator is Hermitian \cite{zwanzig}.

Given an initial density $F(\mathbf{\Gamma },0)$, the formal solution of the
Liouville equation (\ref{liouville}) can be written%
\begin{equation}
F(\mathbf{\Gamma },t)=e^{-itL}F(\mathbf{\Gamma },0).  \label{formal-sol}
\end{equation}%
Since $L$ is Hermitian, all its eigenvalues are real\cite{tolman}, which
according to Eq. (\ref{formal-sol}) implies that $F(\mathbf{\Gamma },t)$
will have an oscillatory behavior. This is an important general result based
on the fully microscopic Hamilton dynamics of the system which allows a
rigorous formulation a closed universe in terms of the Liouville equation (%
\ref{liouville}). A consequence of the Hermitian character of the Liouville
operator is that one may formulate a statistical mechanical theory of a
closed oscillating universe in which the evolution is parametrized by means
of the entropy of the universe itself, and in turn time is parametrized
through the rate of variation of the entropy, i. e., the entropy production.

The paper is organized as follows. In section 2, we formulate the
Generalized Liouville equation and give its formal solution. In section 3,
we postulate the nonequilibrium entropy and obtain the entropy production of
the universe itself. Section 4 is devoted to the derivation of the equations
of the entropic oscillator corresponding to the universe and finally in
section 5 we discuss our main conclusions.

\section{Generalized Liouville equation}

Let us consider that the Hamiltonian of the system is given by 
\begin{equation}
H=\sum_{j=1}^{N}\frac{\mathbf{p}_{j}^{2}}{2m_{j}}+\frac{1}{2}\sum_{j\neq
k=1}^{N}\phi \left( \left\vert \mathbf{q}_{j}-\mathbf{q}_{k}\right\vert
\right) \text{ ,}  \label{hamiltonian}
\end{equation}%
with $m_{j}$ being the mass of a particle, $\mathbf{q}_{j}$ the position
vector of the $j-th$ particle and $\mathbf{p}_{j}$ its conjugated momentum.
Moreover $\phi \left( \left\vert \mathbf{q}_{j}-\mathbf{q}_{k}\right\vert
\right) \equiv \phi _{j,k}$ \ is the interaction potential. The complete
statistical description of the system can be given in terms of the
distribution vector $\mathbf{f}$ \cite{balescu}, \cite{agusti} 
\begin{equation}
\mathbf{f}(t)\equiv \left\{
f_{o},f_{1}(x_{1},t),f_{2}(x^{2},t),.........,f_{N}(x^{N},t)\right\} \text{ }
\label{distribution}
\end{equation}%
which is the set of all the $n$-particle reduced distribution functions,

\begin{equation}
f_{n}=\int F(x^{N},t)\text{ }dx_{n+1}...dx_{N}\text{,}  \label{reduced_dis}
\end{equation}%
where $x_{j}=\left( \mathbf{q}_{j},\mathbf{p}_{j}\right) $ and $%
x^{n}=\left\{ x_{1},...,x_{n}\right\} $, $n=0,......,N$. Here, $f_{o}=1$ and 
$f_{N}(x^{N})=F(x^{N},t)$. The dynamics of the distribution vector $\mathbf{f%
}$ is obtained by integrating the Liouville equation (\ref{liouville})
according to the definition of the $n$-particle reduced distribution
functions (\ref{reduced_dis}), which gives 
\begin{equation}
\frac{\partial }{\partial t}f_{n}=\left\{ H_{n},f_{n}\right\} +\left(
N-n\right) \sum_{j=1}^{n}\int \mathbf{F}_{j,n+1}\frac{\partial }{\partial 
\mathbf{p}_{j}}f_{n+1}dx_{n+1}\text{ .}  \label{n-part_evol}
\end{equation}%
Here, $\mathbf{F}_{j,n+1}=-\mathbf{\nabla }_{j}\phi _{j,n+1}$and 
\begin{equation}
H_{n}=\sum_{i=1}^{n}\frac{\mathbf{p}_{i}^{2}}{2m_{i}}+\frac{1}{2}\sum_{i\neq
k=1}^{n}\phi _{i,k}\text{ }  \label{s-par_hamil}
\end{equation}%
is the $n$-particle Hamiltonian. The first term on the right hand side of
Eq. (\ref{n-part_evol}) represents a Hamiltonian flow, while the second, a
non-Hamiltonian contribution \ due to the coarse graining of the
description. The set of equations represented by Eq. (\ref{n-part_evol})
constitutes a hierarchy of coupled equations, the Bogoliubov, Born, Green,
Kirkwood, Yvon (BBGKY) hierarchy of equations \cite{balescu}, \cite{agusti}
representing the full microscopic description of the system.\textbf{\ }The
BBGKY hierarchy, \ which can be represented in a compact way%
\begin{equation}
i\frac{\partial }{\partial t}\mathbf{f(}t\mathbf{)=}\mathcal{L}\mathbf{f(}t%
\mathbf{),}  \label{gen_lio}
\end{equation}%
constituting\textbf{\ } the Generalized Liouville equation. By comparison of
Eqs. (\ref{n-part_evol}) and (\ref{gen_lio}) it can be seen that the
Generalized Liouville operator $\mathcal{L}$ splits into a diagonal
Hermitian part $\mathcal{PL}$ defined through \cite{balescu}, \cite{agusti} 
\begin{equation}
\langle n\left\vert \mathcal{PL}\right\vert n^{\prime }\rangle =i\left[
H_{n},f_{n}\right] _{P}\delta _{n^{\prime },n}\text{, }n>0\text{ ,}
\label{diago}
\end{equation}%
where $\left\vert n\right\rangle $ represents the $n$-particle state, and a
nondiagonal non-Hermitian part $\mathcal{QL}$ 
\begin{equation}
\left\langle n\right\vert \mathcal{QL}\left\vert n^{\prime }\right\rangle
=i\left\{ \left( N-n\right) \sum_{j=1}^{n}\int \mathbf{F}_{j,n+1}\frac{%
\partial }{\partial \mathbf{p}_{j}}f_{n+1}dx_{n+1}\right\} \text{ }\delta
_{n^{\prime },n+1}\text{ , }\ n>1\text{.}  \label{nodia}
\end{equation}%
Therefore, Eq. (\ref{gen_lio}) can be rewritten%
\begin{equation}
i\frac{\partial }{\partial t}\mathbf{f(}t\mathbf{)}-\mathcal{PL}\mathbf{f(}t%
\mathbf{)=}\mathcal{QL}\mathbf{f(}t\mathbf{)}\text{ }  \label{gen_lio_2}
\end{equation}%
which explicitly manifests the Hermitian and non-Hermitian contributions to
the dynamic of $\mathbf{f(}t\mathbf{)}$. In Eq. (\ref{gen_lio_2}) , the term 
$\mathcal{PL}\mathbf{f(}t\mathbf{)}$ plays the same role as the Liouville
operator in Eq. (\ref{liouville}) and since they are both Hermitian, they
induce comparably incompressible flows. On the other hand, the term $%
\mathcal{QL}\mathbf{f(}t\mathbf{)}$ introduces long range correlations in
the dynamics, which are related to the effects of coarse-grained
interactions, and thus to dissipative effects in a reduced description. This
implies that\textbf{\ }Eq. (\ref{gen_lio_2}) provides us with more
information than Eq. (\ref{liouville}). In fact, as we will see, the
non-Hermitian term is responsible for the approach to equilibrium,
constituting a proof of the adequate description in terms of $\mathbf{f(}t%
\mathbf{)}$ instead of the full phase-space distribution function $F$.

The formal solution of Eq. (\ref{gen_lio_2}) is 
\begin{equation}
\mathbf{f(}t\mathbf{)=}\exp \left( -i\mathcal{PL}t\right) \mathbf{f(}0%
\mathbf{)}+\exp \left( i\mathcal{PL}t\right) \int_{0}^{t}d\tau \exp \left( -i%
\mathcal{PL}\tau \right) (-i\mathcal{QL)}\mathbf{f(}\tau \mathbf{)}\text{ ,}
\label{sol}
\end{equation}%
which, proceeding iteratively, reduces to 
\begin{equation}
\mathbf{f(}t\mathbf{)=}\mathcal{U}\left( t,0\right) \mathbf{f(}0\mathbf{)}%
\text{ \ ,}  \label{pertur_sol}
\end{equation}%
where the evolution operator $\mathcal{U}\left( t,0\right) $ is given by a
perturbative development as 
\begin{gather}
\mathcal{U}\left( t,0\right) \mathbf{=}\sum_{j=0}^{\infty
}\int_{0}^{t}dt_{1}\int_{0}^{t_{1}}dt_{2}\int_{0}^{t_{2}}dt_{3}.......%
\int_{0}^{t_{j-1}}dt_{j}\times  \notag \\
\mathbf{V}\left( t,t_{1}\right) ......\mathbf{V}\left( t_{j-1},t_{j}\right)
\exp \left( -i\mathcal{PL}t_{j}\right) \text{ \ .}  \label{pert_oper}
\end{gather}%
Here, $\mathbf{V}\left( t_{j-1},t_{j}\right) =\exp \left[ i\mathcal{PL}%
\left( t_{j-1}-t_{j}\right) \right] (-i\mathcal{QL)}$ are nonHermitian
propagators, $t_{j}<t_{j-1}<.....<t_{1}<t_{0}=t$, and the integration
proceeds from right to left. Since $\mathcal{PL}$ is Hermitian, all its
eigenvalues are real \cite{tolman}, which means that $\mathbf{f(}t\mathbf{)}$
as given through Eqs. (\ref{pertur_sol}) and (\ref{pert_oper}) will have an
oscillatory behavior. This oscillatory behavior has strong consequences in
the evolution of the closed system, as we will show in the next sections.

\section{Nonequilibrium entropy}

In order to establish a connection between the N-particle microscopic
description given by the generalized Liouville equation (\ref{gen_lio_2})
and a macroscopic coarse-grained description, it is necessary to consider
the statistical definition of the entropy. To achieve this objective, we
start by noticing that the Gibbs entropy 
\begin{equation}
S_{N}=-k_{B}\text{Tr}\left( F\ln F\right) =-k_{B}\int F\ln Fdx^{N}\text{ \ \
,}  \label{Gibbs}
\end{equation}%
where $k_{B}$ is the Boltzmann constant, is a constant of motion under the
Liouville dynamics given through Eq. (\ref{liouville}), hence alternatively,
as the nonequilibrium entropy for the N-body system, we propose \cite{agusti}%
\begin{gather}
S=-k_{B}\text{Tr}\left\{ \mathbf{f}\ln \mathbf{f}_{eq}^{-1}\mathbf{f}%
\right\} +S_{eq}  \notag \\
=-k_{B}\sum_{n=1}^{N}\int f_{n}\ln \frac{f_{n}}{f_{eq,n}}\;dx_{1}.....dx_{n}%
\text{ }+S_{eq}\text{ \ ,}  \label{noneq_entropy}
\end{gather}%
a convex functional of the distribution vector which generalizes the Gibbs
formula. In Eq. (\ref{noneq_entropy}), $S_{eq}$ is the thermodynamic entropy
(\textit{i.e. }the equilibrium entropy) and $\mathbf{f}_{eq}$ is the
equilibrium distribution vector satisfying $\mathcal{L}\mathbf{f}_{eq}=0$,
the Yvon-Born-Green (YBG) equilibrium hierarchy\cite{balescu}. Therefore, $%
\mathbf{f}_{eq}$ is an eigenfunction of $\mathcal{L}$ with eigenvalue $0$.

From the convexity of the logarithmic function, it has been proven \cite%
{grad}, \cite{agusti} that the entropy $S$ varies between the bounds
determined through%
\begin{equation}
0\geq S_{eq}\geq S\geq S_{N}+S_{eq}\text{ .}  \label{bounds}
\end{equation}%
Making use of Eq. (\ref{gen_lio}) we can obtain the rate of change of $S$,
which\textbf{\ }is the entropy production 
\begin{gather}
\frac{\partial S}{\partial t}=ik_{B}\text{Tr}\left\{ \mathcal{L}\mathbf{f}%
\ln \left( \mathbf{f}_{eq}^{-1}\mathbf{f}\right) \right\} =  \notag \\
-\frac{1}{T}\sum_{n=1}^{N}\sum_{j=1}^{n}\int f_{n}\mathbf{p}_{j}\left(
-k_{B}T\frac{\partial }{\partial \mathbf{q}_{j}}\ln f_{eq,n}+\sum_{j\neq
i=1}^{n}\mathbf{F}_{j,i}+\left( N-n\right) \mathcal{F}_{j}\right) dx^{n}%
\text{ .}  \label{entr-pro}
\end{gather}%
Here, $\mathcal{F}_{j}(x^{n})$ is the force on the $j$-th particle from the $%
N-n$\ particles not contained in the cluster of size $n$, and is defined
through the relation:\textbf{\ } $f_{n}(x^{n})$ $\mathcal{F}_{j}(x^{n})=\int 
\mathbf{F}_{j,n+1}f_{n+1}dx_{n+1}$, and $T$ is the kinetic temperature
obtained by\textbf{\ }taking into account that the dependence of $f_{eq,n}$
on the velocities is given through a local Maxwellian. The entropy
production given in Eq. (\ref{entr-pro}) vanishes at equilibrium and in any
other case it should not be necessarily zero. In addition, because $\mathbf{p%
}_{j}$ is arbitrary,%
\begin{equation}
\sum_{j\neq i=1}^{n}\mathbf{F}_{j,i}+\left( N-n\right) \mathcal{F}%
_{j}^{eq}=k_{B}T\frac{\partial }{\partial \mathbf{q}_{j}}\ln f_{eq,n}
\label{YBG}
\end{equation}%
is \textbf{\ }sufficient \textbf{\ }to satisfy the extremum condition $%
\delta \dot{S}/\delta f_{n}\mid _{eq}=0$, with $\dot{S}=\partial S/\partial
t $. Precisely, Eq. (\ref{YBG}) gives rise to the YBG hierarchy\textbf{\ }%
previously mentioned \cite{agusti}$,$\cite{hill}, and expresses a balance of
forces: in the right-hand side of this equation there appear the mean force 
\cite{hill} and in the left-hand side the sum of two terms; the first is the
force due to the van der Waals interactions with the $n-1$ fixed particles
different from the $j$-th particle in the $n$-th cluster (responsible for
the compressions and dilatations of the $n$-th cluster), while the second
which introduces long range correlations is the average force on the $j$-th
particle from the remaining $N-n$ particles of the system.

\section{Entropic oscillator}

In Section 2, we have demonstrated through Eq. (\ref{pertur_sol}) that the
distribution vector $f(t)$\ corresponding to the closed universe possesses
an oscillatory behavior\ and in Section 3, we have established the relation
between $f(t)$\ and the entropy $S$ through Eq. (\ref{noneq_entropy}).
Therefore, as a consequence of these $S(t)$\ will also have an oscillatory
behavior which can be described by%
\begin{equation}
\frac{d}{dt}\left( \frac{\dot{S}(t)}{\sqrt{K_{eff}(t)}}\right) +\sqrt{%
K_{eff}(t)}S=0,  \label{oscillator}
\end{equation}%
if one assumes that $S(t)$ behaves as a free oscillator with an "elastic"
time dependent coefficient $K_{eff}(t)$ defined through the relation%
\begin{equation}
K_{eff}(t)=\frac{1}{S}\frac{\partial ^{2}}{\partial t^{2}}S\text{ .}
\label{elastic_cons}
\end{equation}%
A characteristic value of\textbf{\ } $K_{eff}(t)$ can be estimated from the
consideration that the period of the entropy oscillations of the closed
universe\textbf{\ }should be coherent with the Poincare cycles, $\tau _{p}$,
which leads to 
\begin{equation}
2\pi =\int_{0}^{\tau _{p}}\sqrt{K_{eff}(t)}dt.  \label{period}
\end{equation}%
From these considerations it follows that the entropy of the closed universe
will oscillate between two \textquotedblright equilibrium\textquotedblright\
states:\textbf{\ }$0\geq S_{eq}\geq S\geq S_{N}+S_{eq}$\textbf{. }This is
one of the main results of this paper, since it strongly suggests that our
statistical mechanical theory of a closed universe might complement the
standard cosmological model of an oscillating universe based on purely
general relativity grounds \cite{hawking}.

The implications of this theory on the concept of time can be extracted by
first noticing that the dynamics (\ref{oscillator}) for the entropy of the
universe\textbf{\ }corresponds the Hamiltonian \cite{gentile}%
\begin{equation}
\mathcal{H}\left( S,\dot{S},t\right) =\frac{1}{2}\left( \frac{\dot{S}(t)}{%
\sqrt{K_{eff}(t)}}\right) ^{2}+\Phi (S),  \label{energy}
\end{equation}%
where $\Phi (S)=(1/2)(S-S^{\ast })^{2},$with $S^{\ast }=S_{eq}+S_{N}/2$.
This indicates that the central quantities in the statistical description
are related to the entropy $S$\ and its rate of change $\dot{S}$. Moreover,
since the solution of Eq. (\ref{oscillator}) is given by 
\begin{equation}
S(t)-S^{\ast }=\bigtriangleup _{S}\cos \left[ z(t)+\alpha \right] ,
\label{sol2}
\end{equation}%
where $\bigtriangleup _{S}=S_{N}/2$, and $\alpha $ is a parameter containing
the initial conditions, and where the rescaled time $z(t)$\ is defined
through the relation 
\begin{equation}
z(t)=\int \sqrt{K_{eff}(t^{\prime })}dt^{\prime },  \label{time}
\end{equation}%
the solution given through Eq. (\ref{sol2})\textbf{\ }of the Eq. (\ref%
{oscillator}) enables us to parametrize time in terms of the entropy through
the relation%
\begin{equation}
z(S)=\pm \frac{1}{\sqrt{2}}\int \frac{dt^{\prime }}{\sqrt{\mathcal{H}-\Phi
(S)}}.  \label{parametri}
\end{equation}%
For the potential $\Phi (S)$ already introduced we explicitly obtain $%
z(S)=\pm \sqrt{2}\arcsin \left[ (S-S^{\ast })/\sqrt{2\mathcal{H}}\right] $.
From this relation we may conclude that time depends upon the amount of
entropy generated by the universe during its evolution towards the final
equilibrium state. More precisely, in view of Eq. (\ref{energy}), the
quantity $\sqrt{\mathcal{H}-\Phi (S)}$\ constitutes\textbf{\ } the scaled
entropy production and therefore time in the universe is self-generated by
interactions through the entropy production. From this analysis it follows
that during the evolution of the universe, time is non-homogeneous due to
the non-homogeneous character of the entropy production of the universe
itself.

\section{Conclusions}

In this article, we have proposed a statistical mechanic theory of a closed
universe configured by physical interactions. These interactions define the
lifespan of the universe itself and its volume in the configurational space.
The whole universe can be represented as a point in a two-dimensional phase
space determined by the entropy $S$ and the entropy production $\dot{S}$.
The universe evolves in its phase space along a closed trajectory which
constitutes the phase portrait of a free entropic oscillator. Since the
entropy production depends on the state of the universe through its
instantaneous entropy, we conclude that time has a nonhomogeneous character.

The key point in our \ theory lies in the definition of the state of the
universe by means of the distribution vector and the postulate of the
Generalized Gibbs entropy as a functional of the distribution vector. Unlike
the Gibbs entropy which is a constant of motion under the Liouville
dynamics, the Generalized Gibbs entropy is a varying quantity whose
variation is given by the Generalized Liouville dynamics.

Two main results follow from our analysis: The first one is that our theory
predicts a closed oscillating universe, in accordance with the full
microscopic description given by the Generalized Liouville equation, and
that time is a consequence of dissipative interactions that may be
quantified by the entropy production of the universe itslef.

Quantum and relativistic effects can be taken into account by formulating
our theory in terms of the set of reduced density operators and the
corresponding generalization of \ the von Neumann equation with the suitable
metric, which will be addressed in further works. As it was shown, since the
main results of the present work are a consequence of the nature of the
Liouville eigenvalues, despite of the details of these more convoluted
mathematical formulations, the main conclusion of our model will still be
valid.

\end{document}